\documentclass[prl,twocolumn,amsmath,amssymb,showpacs,aps]{revtex4}
\usepackage{graphicx}

\begin{document}

\title{Spin Pumping in Electrodynamically Coupled Magnon-Photon Systems}

\author{Lihui Bai$^{1}$, M. Harder$^{1}$, Y. P. Chen$^{2}$, X. Fan$^{2,\footnote{Current affiliation: Department of Physics and Astronomy, University of Denver, Colorado, 80208, USA}}$,  J. Q. Xiao$^{2}$, and C.-M. Hu$^{1,\footnote{Electronic address: hu@physics.umanitoba.ca;
URL: http://www.physics.umanitoba.ca/$\sim$hu}}$}

\affiliation{$^{1}$Department of Physics and Astronomy, University
of Manitoba, Winnipeg, Canada R3T 2N2}
\affiliation{$^{2}$Department of Physics and Astronomy, University of Delaware, Newark, Delaware 19716, USA}

\date{\today}

\begin{abstract}
We use electrical detection, in combination with microwave transmission, to investigate both resonant and non-resonant magnon-photon coupling at room temperature. Spin pumping in a dynamically coupled magnon-photon system is found to be distinctly different from previous experiments. Characteristic coupling features such as modes anti-crossing, line width evolution, peculiar line shape, and resonance broadening are systematically measured and consistently analyzed by a theoretical model set on the foundation of classical electrodynamic coupling. Our experimental and theoretical approach pave the way for pursuing microwave coherent manipulation of pure spin current via the combination of spin pumping and magnon-photon coupling.
\end{abstract}

\maketitle

Coupling between electrodynamics and magnetization dynamics is a subject of cross-disciplinary and long-standing interest. The nuclear magnetic resonance (NMR) community has studied this effect for decades by measuring the radiation damping of NMR \cite{Bloemerrgen1954}. Engineers have routinely utilized this effect for designing microwave \cite{Lax1954} and THz devices \cite{THz}. In condensed matter physics, such a coupling leads to the magnon polariton \cite{Mills1974}, which is an elementary excitation characterized by an intrinsic excitation gap between ferromagnetic resonance (FMR) and ferromagnetic antiresonance \cite{Gui2005}. Extrinsically, classical coupling of magnetization dynamics with its electrodynamic surrounding causes Faraday induction of both NMR \cite{Hoult2001} and FMR \cite{Silva1999}. From the perspective of quantum physics, resonant spin-photon coupling plays a central role in utilizing quantum information \cite{Baugh2005}.

In 2010, a theoretical work of Soykal and Flatt$\acute{e}$ \cite{Soykal2010} sparked excitement in the community of spintronics for studying the strong field interaction of magnons and microwave photons. Pioneering experiments have been performed at cryogenic temperatures by Huebl \textit{et al.} \cite{Huebl2013} and Tabuchi \textit{et al.} \cite{Tabuchi2014} on the ferromagnetic insulator Yttrium iron garnet (YIG) placed on/in a microwave cavity, in which a large normal mode splitting was found in the transmission measurements, indicating large quantum-coherent magnon-photon coupling. In October 2014, an experimental breakthrough was made by Zhang \textit{et al.} \cite{Zhang2014}, who demonstrated Rabi-oscillations of the coupled magnon-photon system at room temperature. In the same month, an ultrahigh cooperativity of 10$^5$ between magnon and photon modes was reported \cite{Goryachev2014}. These exciting works reveal just the tip of the iceberg of the new field of cavity spintronics.

So far, experiments in this emerging field were performed by measuring either the transmission ($S_{21}$) or reflection coefficient ($S_{11}$) of the microwave cavity loaded with a YIG sample. The coupling strength was obtained by fitting these $S$ parameters to the microwave input-output formalism with an added self-energy term attributed to the magnon-photon coupling. This standard approach does not specify the underlying coupling mechanism. In this letter, we establish new methods for studying magnon-photon coupling. Experimentally, we demonstrate that spin pumping \cite{Tserkovnyak2005} enables electrical detection of the hybrid magnon-photon modes, showing distinct features not seen in any previous spin pumping experiments. Theoretically, we develop a model based on the combination of a microwave LCR \cite{Chang2005} and Landau Lifshitz Gilbert (LLG) equations, which explains the characteristic features found in our experiment. In addition, we discover the non-resonant magnon-photon coupling via spin pumping measurements, indicating that the standard interpretation of FMR damping might require revisions. Our work sets cavity spintronics on the ground of classical electrodynamic coupling, providing a new perspective for understanding magnon-photon coupling in the general context of dynamic coupling between electro- and magnetization dynamics.

\begin{figure}[t]
\includegraphics[width = 8.5 cm]{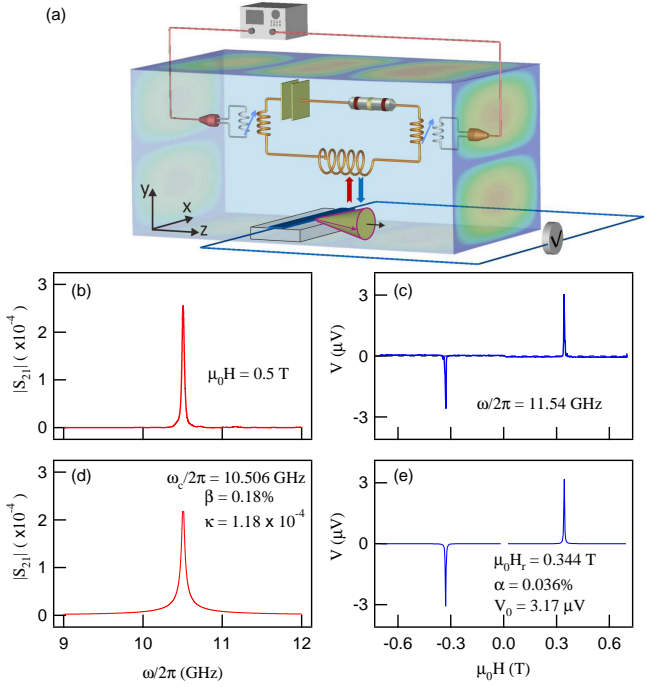}
\caption{ (Color online) (a) Sketch of the experimental setup. The artificial LCR circuit models the microwave current carried by the cavity mode, which couples with the precessing magnetization in YIG. (b) A typical cavity mode measured by transmission spectrum $|S_{21}|$. (c) The uncoupled FMR electrically detected by spin pumping. (d) $|S_{21}|$ and (e) $V(H)$ spectrum calculated by solving Eq. 1 by setting $K_c=K_m$=0.}
\label{sketch}
\end{figure}

Our experimental setup is shown in Fig.\ref{sketch}(a). Comparing with prior experiments \cite{Huebl2013,Tabuchi2014,Zhang2014,Goryachev2014}, it has three special technical features: (i) The microwave cavity made of aluminium has two tuneable connector ports coupled to the input/output microwave circuits. (ii) The empty cavity exhibits several modes in the frequency range of 4 - 16 GHz, with tunable quality factors above 1000. (iii) The setup is designed to enable both transmission measurements of the cavity and the electrical detection of FMR on samples loaded in the cavity. These technical innovations, together with the patterned YIG/platinum(Pt) bilayer sample suitable for spin pumping measurements, reveal unprecedented features of both resonant and non-resonant magnon-photon coupling.

Fig. \ref{sketch}(b) shows a cavity mode at $\omega_c/2 \pi$ = 10.506 GHz measured by $\vert S_{21}\vert$. It's intrinsic loss rate, $\beta = \Delta\omega/\omega_c$ = 0.18\%, is determined from the half width at half maximum (HWHM) $\Delta\omega$. The YIG(2.6 $\mu$m)/Pt(10 nm) bilayer sample is characterized before loading into the cavity by sending a microwave current at $\omega/2 \pi$ =  11.54 GHz into the Pt layer which is patterned on top of the YIG film into a strip with a dimension of 50 $\mu$m $\times$ 5 mm. By modulating the microwave current at 8.33 kHz to use the lock-in technique, a dc voltage $V(H)$ is measured along the Pt strip as a function of the external magnetic field $\mathbf{H}$ that is applied in the sample plane but perpendicular to the Pt strip. We use the unit vectors $\textbf{e}_{x}$, $\textbf{e}_{y}$ and $\textbf{e}_{H}$ to denote the long axis of the Pt strip, sample normal, and the $\mathbf{H}$ field direction. The sharp resonances appearing at $\mu_0 H_r$ = $\pm$ 0.344 T in Fig. \ref{sketch}(c) have the dispersion relation  of $\omega$ = $\gamma \sqrt{|H_r|(|H_r|+M_0)}$ (plotted in Figs. 2(c) and (d) as dashed lines), which we identify as the YIG FMR. Here, $\gamma = 172.7 \mu_{0}$ GHz/T and $\mu_{0}M_{0}$ = 0.169 T are the gyromagnetic ratio and the saturation magnetization of YIG respectively. Using the well-known relation \cite{Sparks1964,Heinrich1985} of $\Delta H = \Delta H_0 + \alpha\omega/\gamma$ to fit the HWHM $\Delta H$ measured at different frequencies, we determine the intrinsic Gilbert damping parameter $\alpha$ = 0.036\% and the zero-frequency intercept $\mu_0\Delta H_0$ = 0.31 mT. Note that studying damping mechanisms is a core subject of magnetism \cite{Tserkovnyak2005,Sparks1964,Heinrich1985}, and it is generally accepted that $\Delta H_0$ is sample-specific but independent of the microwave fields. Also, it is generally accepted that FMR measured by spin pumping has a characteristic Lorentz line shape \cite{Tserkovnyak2005,Bai2013,Sinova2014}, as shown in Fig. \ref{sketch}(c). These views will be revised by our experiment studying magnon-photon coupling.

\begin{figure}[!t]
\includegraphics[width = 8.5 cm]{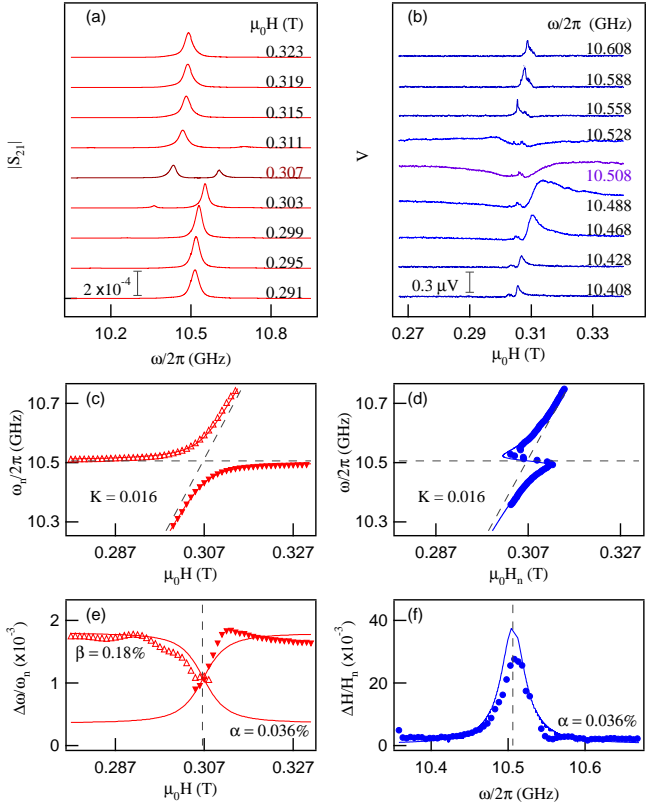}
\caption{ (Color online) Evolution of the hybrid FMR and cavity mode measured by (a) transmission spectra $|S_{21}|$  and (b) spin pumping voltage $V(H)$. Dispersion of the hybrid modes determined from (c) $|S_{21}|$ and (d) $V(H)$ spectra. Normalized line width of the hybrid modes determined from (e) $|S_{21}|$ and (f) $V(H)$ spectra. The solid curves in (c)-(f) are calculated by solving Eq. 1.}
\label{rawdata}
\end{figure}

The coupling is achieved by tuning the H-field to let the FMR approach the cavity mode. Mode hybridization is measured in $|S_{21}(\omega)|$ by sweeping $\omega$ at fixed magnetic fields as shown in Fig. \ref{rawdata}(a). The evolution of the cavity-like mode $\omega_n$ ($n$ =1,2) is consistent with prior experiments \cite{Huebl2013,Tabuchi2014,Zhang2014,Goryachev2014}. The new experiment that we conduct here is the spin pumping of YIG FMR driven by the cavity mode, performed by measuring $V(H)$ at fixed microwave frequencies.

The electrically detected FMR-like mode appearing at $H_n$ ($n$=1) shown in Fig. \ref{rawdata}(b) looks rather different than the cavity-like mode plotted in Fig. \ref{rawdata}(a). Their different mode dispersions are plotted in Figs. \ref{rawdata}(c) and (d) for comparison. The contrast is more pronounced if we compare the evolution of the normalized line widths, $\Delta\omega/\omega_n$ and $\Delta H/ H_n$ plotted in Figs. \ref{rawdata}(e) and (f) respectively. Measured by $|S_{21}(\omega)|$, $\Delta\omega$ decreases when approaching the resonant coupling condition at $\mu_0 H$ = 0.307 T. In contrast, measured by $V(H)$, $\Delta H$ increases drastically when approaching the resonant coupling condition at $\omega/2 \pi$ = 10.506 GHz. Furthermore, accompanied with the resonance broadening, a remarkable feature in Fig. \ref{rawdata}(b) is that when the FMR-like mode approaches the resonant coupling condition, it's line shape becomes increasingly asymmetric. This is peculiarly different from the FMR line shape studied in any previous spin pumping experiments, including that shown in Fig. \ref{sketch}(c).

Intrigued by the seminal work of Bloembergen and Pound published in 1954 studying NMR \cite{Bloemerrgen1954}, we develop a concise model set on the footing of mutual coupling between electro and magnetization dynamics to analyze the experimental features. Without coupling, in the rotational frame of the magnetization precession, the FMR and its phase function are determined by the LLG equation of dynamic magnetization $m^+(t) \equiv m_x(\omega_0/\omega_r) + im_y = m e^{-i\omega t}$, where $\omega_0$ = $\gamma |H|$ and $\omega_r = \gamma \sqrt{|H|(|H|+M_0)}$. In the same rotational frame, the cavity resonance can be modeled by an artificial LCR circuit with inductor $L$, capacitor $C$, and resistor $R$, which carries the microwave current $j^+(t) \equiv j_x(\omega_0/\omega_r) + ij_y = j e^{-i\omega t}$. The LCR equation \cite{Chang2005} determines the cavity mode and its phase function. With magnon-photon coupling, two coupling parameters $K_c$ and $K_m$ are introduced in two sets of coupling terms. One set follows the electrodynamic phase relation determined by Faraday$\textquoteright$s law, inducing voltages $V_x(t) = K_cL(dm_y/dt)$ and $V_y(t) = -K_cL(dm_x/dt)$ in the LCR circuit. This is known as the Faraday induction of FMR \cite{Silva1999}. The other set follows the phase relation determined by Amp$\grave{e}$re$\textquoteright$s law, producing magnetic fields $h_x(t) = K_mj_y(t)$ and $h_y(t) = -K_mj_x(t)$ that place a torque on the YIG magnetization. Our simple approach neglects the tedious geometric details. Instead, it \textit{accurately} models and highlights the key physics of phase correlation between $m^+(t)$ and $j^+(t)$ governed by electrodynamics. It is straightforward to prove that our approach leads to the following eigenvalue equations:
\begin{equation}
\begin{pmatrix} \omega^2-\omega_c^2+i2\beta\omega_c\omega & i\omega^2K_c\\-i\omega_mK_m & \omega-\omega_r+i\alpha\omega \end{pmatrix} \begin{pmatrix} j \\ m \end{pmatrix} = 0,
\label{coupling}
\end{equation}
where $\omega_m$ = $\gamma M_0$, and $\omega_c = 1/\sqrt{LC}$.

Setting $K_c = K_m = 0$, Eq. \ref{coupling} shows that the spectral function for the uncoupled cavity mode and FMR is given by $\mathbb{S}_c(\omega) = 1/ (\omega^2-\omega_c^2+i2\beta\omega_c\omega)$ and $\mathbb{S}_m(\omega) = 1/(\omega-\omega_r+i\alpha\omega)$, respectively. Using the microwave input-output formalism \cite{Chang2005} and spin pumping theory \cite{Tserkovnyak2005,Bai2013} in the limit of small external loss and small precession angle, respectively, we obtain $|S_{21}| \approx  2\kappa\sqrt{2\beta\omega\omega_c|\mathrm{Im}(\mathbb{S}_c)|}$ and $V_{\mathrm{SP}} = V_0 \alpha\omega|\mathrm{Im}(\mathbb{S}_m)|~\textbf{e}_{x}\cdot(\textbf{e}_{y}\times\textbf{e}_{H})$ for calculating the transmission coefficient and spin pumping voltage. Here $\kappa \ll 1$ is the external loss rate of the cavity and $V_0$ is the maximum spin pumping voltage that depends on the sample, cavity mode, and microwave power. As shown in Figs. \ref{sketch}(d) and (e), the calculated spectra using $\kappa$ = 1.18$\times$10$^{-4}$ and $V_0$ = 3.17 $\mu$V characterize the experimental curves plotted in Figs. \ref{sketch}(b) and (c) very well.

In the linear coupling regime, both $K_c$ and $K_m$ are nonzero but they are linked by the Onsager reciprocal relations. Hence, the 2 $\times$ 2 eigenvalue problem defined by Eq. \ref{coupling} is determined by only one dimensionless coupling constant $K \equiv \sqrt{K_cK_m}$.

By solving the complex eigenfrequencies $\tilde{\omega}_n$ ($n$ = 1,2) of Eq. \ref{coupling} at fixed H-fields, we plot in Figs. \ref{rawdata}(c) and (e), respectively, the resonance frequency Re($\tilde{\omega}_n$) and the normalized line width $|$Im($\tilde{\omega}_n$)$|$/Re($\tilde{\omega}_n$) of the hybrid modes calculated by setting $K$ = 0.016.  The calculated dispersion plotted in Figs. \ref{rawdata}(c) agrees very well with the measured data. Accompanied with the evolution of the cavity mode towards FMR, $|$Im($\tilde{\omega}_n$)$|$/Re($\tilde{\omega}_n$) decreases from $\beta$ = 0.18\%  to $\alpha$ = 0.036\%. This explains why $\Delta\omega/\omega_n$ decreases in Fig. \ref{rawdata}(e) at the resonant coupling condition \cite{HOM}.  Note that if $\alpha > \beta$, our model gives the opposite result (not shown) of $|$Im($\tilde{\omega}_n$)$|$/Re($\tilde{\omega}_n$) increasing from $\beta$ to $\alpha$, which explains the transmission spectra measured by Huebl \textit{et al.} \cite{Huebl2013}. We also predict that in the case of $\alpha = \beta$, no line width change will be observed in transmission/reflection measurements. In the following, we further demonstrate that the drastically different appearance of the spin pumping spectra is also governed by the same physics described by Eq. \ref{coupling}.

In Figs. \ref{rawdata}(d) and (f), we plot the results calculated by finding the complex eigenvalue $\tilde{H_n}$ of Eq. \ref{coupling} at fixed microwave frequencies. By using the same coupling constant of $K$ = 0.016, both the resonance field Re($\tilde{H_n}$) and the normalized line width $|$Im($\tilde{H_n}$)$|$/Re($\tilde{H_n}$) agree with the spin pumping measurements. It shows that although $\alpha$ is not changed by the coupling, $\Delta H$ increases drastically at the resonant coupling condition. Remarkably, in contrast to $|$Im($\tilde{\omega}_n$)$|$/Re($\tilde{\omega}_n$) discussed above, we find that $|$Im($\tilde{H_n}$)$|$/Re($\tilde{H_n}$) always increases as the FMR approaches the resonant coupling condition, no matter whether $\alpha < \beta$ or $\alpha \ge \beta$. Such a general feature implies that the standard relation \cite{Sparks1964} of $\Delta H = \Delta H_0 + \alpha\omega/\gamma$ is no longer valid for FMR coupled to the cavity mode. This observation leads us to discover the non-resonant magnon-photon coupling, which we will discuss after briefly addressing the peculiar FMR line shape.

Apparently, both the anomalous dispersion and the increased line width are associated with the peculiar line shape shown in Fig. \ref{rawdata}(b). Note that $V_{\mathrm{SP}} \propto|m|^2 \propto |\mathbb{S}|^2$ \cite{Bai2013}, where the spectral function $\mathbb{S}$ for the eigenmode of Eq. \ref{coupling} depends on the coupling. Without coupling, it is easy to prove that $|\mathbb{S}|^2 \propto \mathrm{Im}(\mathbb{S}_m) \propto (\Delta H)^{2}/[(H-H_{r})^2 + (\Delta H)^2]$. That's why spin pumping of uncoupled FMR always has a Lorentz line shape. With coupling, Eq. \ref{coupling} shows that $j$ and $m$ are coherently mixed. Hence, the peculiar line shape of the coupled mode is caused by the phase correlation of $j^+(t)$ and $m^+(t)$, similar to the case of spin rectification \cite{Gui2007}, where $j^+(t)$ and $m^+(t)$ are coherently mixed by the anisotropic magnetoresistance. Indeed, the peculiar line shapes in both cases can be fit by using the combination of symmetric and anti-symmetric Lorentzian functions \cite{Harder2011}. The fact that such a phase correlation effect is now measured by spin pumping is of particular interest, since it implies that the line shape of spin pumping FMR can be tuned by changing the phase difference between $j^+(t)$ and $m^+(t)$, as was demonstrated for spin rectification \cite{Wirthmann2010}. This may enable microwave coherent control of the pure spin current produced by spin pumping, which was celebrated in semiconductor spintronics by using coherent optical methods \cite{Zhao2006,Werake2010}. Here, we briefly demonstrate the feasibility of line shape tuning, before elucidating the significant effect of non-resonant coupling, which has so-far escaped prior investigations \cite{Soykal2010,Huebl2013,Tabuchi2014,Zhang2014,Goryachev2014}.

\begin{figure}[!t]
\includegraphics[width = 8.5 cm]{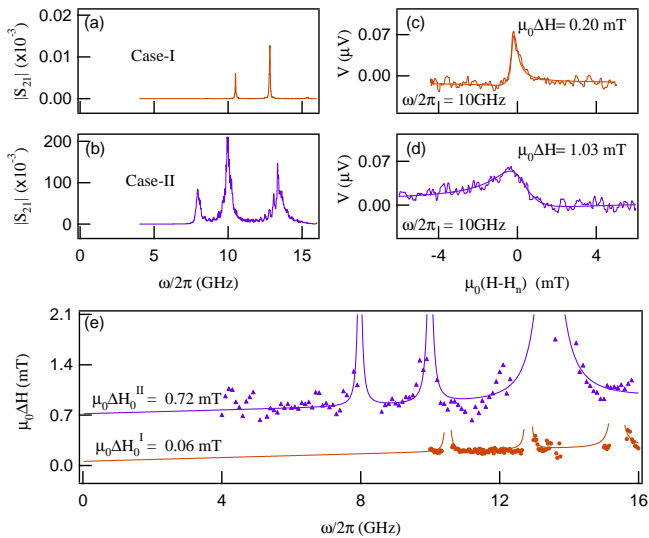}
\caption{ (Color online) $|S_{21}|$ spectra measured (a) before and (b) after tuning the adjustable connector port of the cavity. The FMR line shape measured (c) before and (d) after tuning the connector port. (e) Comparison of the frequency dependence of the FMR line width measured before and after tuning the connector port. Solid curves are fits to Eq. 2.}
\label{Linewidth_S21}
\end{figure}

Using the adjustable connector port between the cavity and its input microwave circuit, we can control the cavity modes and their coupling to the FMR. Two cases are compared here by only tuning a connector port  \cite{Nonlinear}. In case-I, three cavity modes of $(\omega_c/2\pi [\textrm{GHz}], \beta)$ = (10.502, 0.12\%), (12.822, 0.12\%), (15.362, 0.35\%) are excited, as shown in Fig. \ref{Linewidth_S21}(a). In case-II shown in Fig. \ref{Linewidth_S21}(b), three modes of (7.969, 0.79\%), (9.990, 0.88\%) and (13.414, 2.38\%) are excited. Accordingly, as found in Figs. \ref{Linewidth_S21}(c) and (d), the line shape of the spin pumping spectrum $V(H)$ measured at $\omega/2\pi$ = 10 GHz is significantly tuned by the microwave. Accompanied with the tuned line shape is the increased $\Delta H$, obtained by fitting $V(H)$ to the asymmetric Lorentz line shape function \cite{Harder2011}. Using our special set up, we can study $\Delta H(\omega)$ in a broad band that covers non-resonant regions between the resonant cavity modes. It is in such a measurement that the surprising effect of non-resonant magnon-photon coupling is revealed.

In Fig. \ref{Linewidth_S21}(e), we plot and compare $\Delta H(\omega)$ measured in both cases, and we fit the data to the equation
\begin{equation}
\Delta H(\omega) = \Delta H_{0} + \frac{\alpha\omega}{\gamma} + \frac{\omega^{2}\omega_{m}}{\gamma}\sum_{l}K_{l}^{2}|\mathrm{Im}(\mathbb{S}_{c,l})|.
\label{linewidth}
\end{equation}
The last term in Eq. 2 follows from Eq. 1, which describes the coupling-enhanced FMR line width near each cavity mode \cite{fit}. In case-I, we obtain $\alpha$ = 0.036\% and $\Delta H_{0}^I$ = 0.06 mT/$\mu_0$ from the fitting. As mentioned, $\Delta H_0$ is generally attributed to sample-specific FMR damping. Hence, it is surprising that $\Delta H_{0}^I \ll \Delta H_{0}$ = 0.31 mT/$\mu_0$ that was characterized before loading the sample into the cavity. We note that $|S_{21}|$ plotted in Fig. \ref{Linewidth_S21}(a) shows that there are very few cavity states carrying microwaves in the non-resonant regions. It suggests that the reduced $\Delta H_{0}^I$ may be related to the suppressed microwave density of states \cite{Lagendijk1996,Beenakker1991} in the cavity.

This is confirmed by tuning the cavity to case-II. Here, as shown in Fig. \ref{Linewidth_S21}(b), $|S_{21}|$ is significantly enhanced. Accordingly, the measured $\Delta H$ plotted in Fig. \ref{Linewidth_S21}(e) is increased not only near each cavity mode, but also in the non-resonant regions between these modes. From the fitting, we obtain the same $\alpha$ = 0.036\%, but now we get $\Delta H_{0}^{II}$ = 0.72 mT/$\mu_0$ that is more than one order of magnitude larger than $\Delta H_{0}^I$ = 0.06 mT/$\mu_0$. This unveils the significant effect of FMR broadening due to non-resonant magnon-photon coupling. Such an extrinsic damping caused by the coherent coupling of FMR with microwave fields is even larger than the intrinsic Gilbert damping in YIG determined by the spin-spin and the spin-lattice relaxation mechanisms.

In summary, by using a combination of transmission and electrical detection methods, and by developing a concise model on the general footing of electrodynamic coupling, we have studied the signatures of resonant magnon-photon coupling, showing that the underlying key physics is the phase correlation of magnetization and electro-dynamics \cite{SM}. The scope of our theoretical formalism can be easily extended to involve magnon polariton and spin waves \cite{SM}. In addition to the resonant coupling, our experiment also unveils the effect of non-resonant magnon-photon coupling, which revises the current understanding of FMR damping. We believe that the experimental and theoretical approaches we report here, with versatile capability and easy accessibility respectively, will make possible detailed studies of cavity spintronics, and pave new ways for controlling spin current via spin pumping and magnon-photon coupling.

Work in Manitoba (experiment and theory) was funded by NSERC, CFI, and URGP grants (C.-M.H.). Work in Delaware (bilayer sample fabrication) was supported by the Semiconductor Research Corporation through the Center for Nanoferroic Devices. The Manitoba-Delaware collaboration was promoted by the Overseas Chinese Scholars Innovation Team led by X.F. Han. We thank B.M. Yao, Z.H. Zhang, and Y. Huo for helps, G. Bridges, H. Guo, Y. Xiao, T. Silva, and S. Goennenwein for discussions.



\begin{thebibliography}{20}

\bibitem{Bloemerrgen1954}
N. Bloembergen and R.V. Pound, Phy. Rev. {\bf 95}, 8 (1954).

\bibitem{Lax1954}
B. Lax, K.J. Button, and L.M. Roth, J. Appl. Phy. {\bf 25}, 1413 (1954).

\bibitem{THz}
K.J. Chau, Mark Johnson, and A.Y. Elezzabi, Phy. Rev. Lett. {\bf 98}, 133901 (2007).

\bibitem{Mills1974}
D.L. Mills, and E. Burstein, Rep. Prog. Phys. {\bf 37}, 817 (1974).

\bibitem{Gui2005}
Y.S. Gui, S. Holland, N. Mecking, and C.-M. Hu, Phy. Rev. Lett. {\bf 95}, 056807 (2005).

\bibitem{Hoult2001}
D.I. Hoult, and N.S. Ginsberg, J. Magn. Resonance. {\bf 148}, 182 (2001).

\bibitem{Silva1999}
T.J. Silva, C.S. Lee, T.M. Crawford, and C.T. Rogers, J. Appl. Phys. {\bf 85}, 7849 (1999).

\bibitem{Baugh2005}
J. Baugh, O. Moussa, C.A. Ryan, A. Nayak, and R. Laflamme, Nature {\bf 438}, 470 (2005).

\bibitem{Soykal2010}
\"{O}.O. Soykal, and M.E. Flatt$\acute{e}$, Phy. Rev. Lett. {\bf 104}, 077202 (2010).

\bibitem{Huebl2013}
H. Huebl, C.W. Zollitsch, J. Lotze, F. Hocke, M. Greifenstein, A. Marx, R. Gross, and S.T.B. Goennenwein, Phy. Rev. Lett. {\bf 111}, 127003 (2013).

\bibitem{Tabuchi2014}
Y. Tabuchi, S. Ishino, T. Ishikawa, R. Yamazaki, K. Usami, and Y. Nakamura, Phy. Rev. Lett. {\bf 113}, 083603 (2014).

\bibitem{Zhang2014}
X. Zhang, C.-L. Zou, L. Jiang, and H.X. Tang, Phy. Rev. Lett. {\bf 113}, 156401 (2014).

\bibitem{Goryachev2014}
M. Goryachev, W. G. Farr, D. L. Creedon, Y. Fan, M. Kostylev, and M. E. Tobar, Phy. Rev. A {\bf 2}, 054002 (2014).

\bibitem{Tserkovnyak2005}
Y. Tserkovnyak, A. Brataas, G.E.W. Bauer, and B. Halperin, Rev. Mod. Phys. {\bf 77}, 1375 (2005).

\bibitem{Chang2005}
A. Luiten, in \textit{Encyclopedia of RF and Microwave Engineering}, edited by Kai Chang (Wiley, Hoboken, N.J., 2005), Vol. 5.

\bibitem{Sparks1964}
M. Sparks, \textit{Ferromagnetic-Relaxation Theory} (McGraw-Hill, New York, 1964).

\bibitem{Heinrich1985}
B. Heinrich, J. F. Cochran, and R. Hasegawa, J. Appl. Phys. {\bf 57}, 3690 (1985).

\bibitem{Bai2013}
L.H. Bai, P. Hyde, Y.S. Gui, C.-M. Hu, V. Vlaminck, J.E. Pearson, S.D. Bader, and A. Hoffmann, Phy. Rev. Lett. {\bf 111}, 217602 (2013).

\bibitem{Sinova2014}
J. Sinova, S.O. Valenzuela, J. Wunderlich, C.H. Back, and T. Jungwirth, arXiv:1411.3249 (2014).

\bibitem{HOM}
Fine structures neglected in our calculations are caused by higher-order coupling involving more than one cavity mode.

\bibitem{Gui2007}
Y.S. Gui, N. Mecking, X. Zhou, G. Williams, and C.-M. Hu, Phys. Rev. Lett. {\bf 98}, 107602 (2007); N. Mecking, Y.S. Gui, and C.-M. Hu, Phys. Rev. B {\bf 76}, 224430 (2007).

\bibitem{Harder2011}
M. Harder, Z.X. Cao, Y.S. Gui, X. L. Fan, and C.-M. Hu, Phys. Rev. B {\bf 84}, 054423 (2011).

\bibitem{Wirthmann2010}
A. Wirthmann, X. Fan, Y. S. Gui, K. Martens, G. Williams, J. Dietrich, G. E. Bridges, and C.-M. Hu, Phys. Rev. Lett. {\bf 105}, 017202 (2010).

\bibitem{Zhao2006}
H. Zhao, E.J. Loren, H.M. van Driel, and A.L. Smirl, Phys. Rev. Lett. {\bf 96}, 246601 (2006).

\bibitem{Werake2010}
L.K. Werake and H. Zhao, Nat. Phys. {\bf 6}, 875 (2010).

\bibitem{Nonlinear}
Precaution has been taken to avoid nonlinear FMR effects, by ensuring a very small precession angle of FMR.

\bibitem{fit}
We set $K_l$ = 0.016 ($l$ = 1,2,3) except for the 3rd mode at 13.414 GHz with $\beta$=2.38\% in case-II, for which we use $K_3$=0.032. The anomalously large $K$ and $\beta$ for this mode indicates that it may involve more than one cavity mode.

\bibitem{Lagendijk1996}
A. Lagendijk and B.A. van Tiggelen, Physics Reports {\bf 270}, 143 (1996).

\bibitem{Beenakker1991}
C.W.J. Beenakker and H. van Houten, Quantum Transport in Semiconductor Nanostructures,
in: \textit{Solid State Physics}, editors H. Ehrenreich and D. Tumbull (Academic, New York, 1991).

\bibitem{SM}
See Supplemental Material at http://link.aps.org/supplemental for more details about the physics of coupling-induced phase correlation and magnon polariton, the experimental method of tuning the microwave cavity, and additonal data analysis involving spin waves.


\end{thebibliography}
\end{document}